# The Fermilab Test Beam Facility Data Acquisition System Based on *otsdaq*

Kurt Biery, Eric Flumerfelt, Adam Lyon, Ron Rechenmacher, Ryan Rivera, Mandy Rominsky, Lorenzo Uplegger, Margaret Votava

*Abstract* – The Real-Time Systems Engineering Department of the Scientific Computing Division at Fermilab has deployed set of customizations to our Off-The-Shelf Data Acquisition solution (*otsdaq*) at the Fermilab Test Beam Facility (FTBF) to read out the beamline instrumentation in support of FTBF users. In addition to reading out several detectors which use various detection technologies and readout hardware, the FTBF Data Acquisition system (DAQ) can perform basic track reconstruction through the facility in real time and provide data to facility users. An advanced prototype smart coincidence module, known as the NIM+, performs trigger distribution and throttling, allowing the beamline instrumentation to be read out at different rates. Spill data are saved to disk for studies of the facility performance, and hit data are also made available on the FTBF network for experiments' use. A web-based run control and configuration GUI are provided, and the online monitoring snapshots created after each beam spill are viewable from any computer connected to the Fermilab network. The integrated DAQ system for the facility provides users with high-precision tracking data along the beamline and a single location for obtaining data from the facility detectors, which set the baseline for testing their own detectors.

## I. Introduction

The Fermilab Test Beam Facility (FTBF) is a high-energy test beam used for precision tests of high-energy physics (HEP) detectors [1]. Customers include large experiments such as CMS [2] and NOvA [3], and smaller research groups testing novel detector technologies. The beamline delivers a user-selected intensity, up to one million particles per four-second spill. Additionally, a range of particle types and energies can be delivered, including 120 GeV protons, based on users' needs.

FTBF's beamline is equipped with several detectors for beam position and energy measurement (Fig. 1). Depending on a specific user's needs and the physical limitations of the space, they may place their experimental setup anywhere along the beamline and several of the beamline detectors may be moved to accommodate them. Experiments using the facility often use position data obtained from the facility's detectors in their analysis, even if only for beam presence verification.

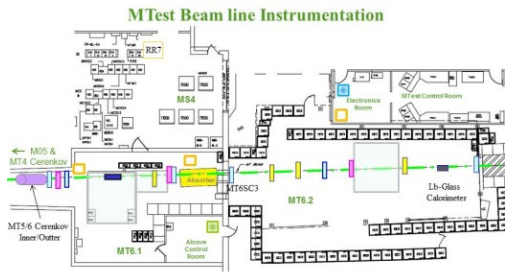

Fig. 1. Diagram of the FTBF beamline

The Real-Time Systems Engineering department within the Fermilab Scientific Computing Division has developed the *otsdaq* product as a part of the *artdaq* toolkit [4]. *otsdaq* is a "ready-to-use" Data Acquisition (DAQ) system, needing very little customization for simple installations. Implementing a unified Facility DAQ with *otsdaq* showcases this product to the test beam experiments, who may then choose to use it for their own DAQ needs. The flexibility of the *artdaq* toolkit used by *otsdaq* allows it to fulfill the needs of users from bench-top experimenters to large detector DAQs. The FTBF DAQ includes several custom components due to its dual nature as a full-featured DAQ and a provider of data to other (possibly non-*artdaq*) DAQs.

## II. The OTSDAQ DAQ Solution

The *artdaq* toolkit provides many options to customize virtually every aspect of a DAQ system. As a part of its mission to create an "off the shelf" DAQ, *otsdaq* makes many of these choices on the behalf of its users. A web-based Run Control GUI (Fig. 2) client was developed that interfaces with the *otsdaq* server built using the XDAQ toolkit from CERN [5], and *otsdaq* controls the *artdaq* processes comprising the DAQ using SOAP messaging. *otsdaq* also provides a configuration editor which communicates with a document-based database through the *artdaq_database* package.

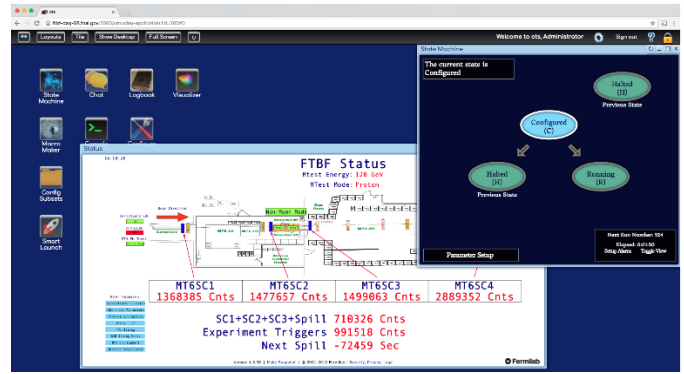

Fig. 2. The *otsdaq* Web Desktop

*artdaq* is used for filtering, analysis, and data transport in the *otsdaq* system. It also provides scalability and modularity, allowing components to be added and removed in the configuration between runs. Data analysis is implemented with the *art* framework [6], and user-defined filtering and analysis modules can be inserted in the data path as desired. Results of the online processing are stored in the data stream, and can be reused in offline processing.

*otsdaq* also provides tools useful during front-end development, calibration, and debugging: the MacroMaker tool allows user interaction with hardware in a simple GUI interface, and a simple data reader allows saving of raw binary data to disk to aid in detector performance debugging. *otsdaq* is very component-oriented, and the Smart Launch utility allows for runtime components to be selected from the main Web GUI desktop. The Iterator tool allows for automatically iterating over a parameter space to efficiently evaluate the detector response. An integrated real-time ROOT viewer, built using JSROOT, allows users to monitor their DAQ with *art* modules which create ROOT objects.

The *otsdaq* web tools that are distributed with the core packages are example applications built on the underlying JavaScript *otsdaq* application programming interface (API). A "Template Supervisor" is provided as a starting point for client application development. Each user is encouraged to create their own set of repositories to manage their client and server-side plugin code, customized to their needs. The *otsdaq* web desktop can be configured, based on user access permissions, to include or omit desktop icons to core and user web applications allowing for a completely custom user experience for each user of the DAQ system.

### III. IMPLEMENTING THE FACILITY DAQ

Prior to this work, each detector in the FTBF beamline had a separate DAQ system. The CMS Strip Telescope was already using *otsdaq*, so it formed the basis of the new system. The first external detector to be integrated was the Multi-Wire Proportional Counters (MWPCs). Because of limitations in the MWPC front-end controller, however, the readout of these detectors cannot be triggered more than ~10,000 times per spill. To overcome this limitation, a Fermilab-developed smart coincidence module (NIM+) was added to the FTBF DAQ to throttle the MWPC trigger rate relative to that of the strip telescope, and allow the Facility DAQ to sample the MWPCs accurately as the beam intensity increases.

The FTBF detectors read out at different rates, so the beam spill was chosen as the "event" for the DAQ. The NIM+ smart coincidence module from Fermilab's Physics Research Equipment Pool (PREP) was used to generate and distribute trigger signals to the different detectors. Data from the NIM+ module allow for trigger-by-trigger matching and breaking the spill down into its component trigger events (corresponding to a particle through the facility).

Several *art* data analysis modules were written for the FTBF DAQ to perform online monitoring on both a per-detector and facility-wide basis. Rough tracking is done through the facility (fine tracking requires more intensive, offline analysis) and hit maps are shown for each detector (Fig. 3).

Since it is a user facility, the Facility DAQ must be able to provide data to other DAQ systems. This can take many forms, from simple data-file handoffs to real-time event serving. An *art* module was written to stream serialized ROOT [7] objects to requestors which connect using a well-defined request command to an *artdaq* Dispatcher running as part of the Facility DAQ.

### IV. CONCLUSION

The FTBF DAQ is an interesting DAQ case study for several reasons. It spans multiple different detector technologies across a large physical area, each of which uses different readout technologies. It also acts as a front-end for other DAQ systems in addition to its primary responsibility of collecting and writing facility data. Finally, it serves as a demonstration of the capabilities and use of the *otsdaq* product and the NIM+ module in a place where it is highly visible to potential users.

A single DAQ system based on *otsdaq* for the facility allows for more precise beamline measurements to be made, and allows users to obtain data from the facility from a centralized location. Having data from the entire beamline allows for more precise tracking and energy estimation, and the facility detectors serve as cross-checks on each other's performance, allowing users to be confident in the baseline thus established. Precision studies of test detectors are possible by identifying the exact location of each individual particle passing through them.

### REFERENCES

[1] "Fermilab Test Beam Facility," [Online]. Available: https://ftbf.fnal.gov.

[2] S. Chatrchyan et al., "The CMS Experiment at the CERN LHC," *JINST,* vol. 3, p. S08004, 2008.

[3] NOvA Collaboration, "The NOvA Technical Design Report," 2007.

[4] K. Biery, C. Green, J. Kowalkowski, M. Paterno and R. Rechenmacher, "artdaq: An Event-Building, Filtering, and Processing Framework," *IEEE Trans. Nucl. Sci.,* vol. 60, pp. 3764-3771, 2013.

[5] CMS Collaboration, "The CMS data acquisition system software," *Journal of Physics: Conference Series,* vol. 219, p. 022011, 4 2010.

[6] C. Green, J. Kowalkowski, M. Paterno, M. Fischler, L. Garren and Q. Lu, "The Art Framework," *J. Phys. Conf. Ser.,* vol. 396, p. 022020, 2012.

[7] R. Brun and F. Rademakers, "ROOT: An object oriented data analysis framework," *Nuclear Instruments and Methods in Physics Research Section A: Accelerators, Spectrometers, Detectors and Associated Equipment,* vol. 389, pp. 81-86, 1997.

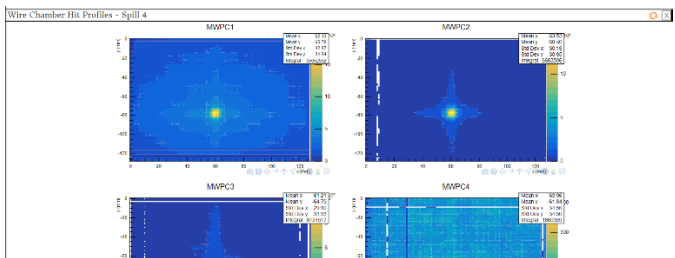

Fig. 3. Data Quality plot for the MWPC detectors. Beam spot is clearly visible. Y-axis is reversed to reflect physical arrangement of detectors. Axes are in millimeters.